\documentclass[twocolumn,showkeys,aps,prd,10pt]{revtex4-1}

\usepackage{amsfonts,amsmath,amssymb}
\usepackage{graphicx,graphics}
\usepackage{epstopdf}
\usepackage{epsfig}
\usepackage{natbib}

\def\mnras{MNRAS}
\def\apj{ApJ}
\def\apjl{ApJL}

\def\prl{Phys. Rev. Lett.}

\begin{document}

\title{What can we really infer from GW 150914?}

\author{J.~F.~Rodriguez,$^{1,2}$ J.~A.~Rueda,$^{1,2,3}$ R.~Ruffini$^{1,2,3}$}
\affiliation{$^1$Dipartimento di Fisica and ICRA, Sapienza Universit\`a di Roma, P.le Aldo Moro 5, I--00185 Rome, Italy}
\affiliation{$^2$ICRANet, Piazza della Repubblica 10, I--65122 Pescara, Italy}
\affiliation{$^3$ICRANet-Rio, CBPF, Rua Dr. Xavier Sigaud 150, Rio de Janeiro, RJ, 22290--180, Brazil}
\date{\today}

\begin{abstract}
In a recent letter we have outlined some issues on GW 150914, we hereby give additional details. We analyze the event GW 150914 announced by the Advanced Laser Interferometer Gravitational-Wave Observatory (LIGO) as the gravitational-wave emission of a black-hole binary merger. We show that the parameters of the coalescing system and of the newly formed Kerr black-hole can be extracted from basic results of the gravitational-wave emission during the inspiraling and merger phases without sophisticate numerical simulations. Our strikingly accurate estimates are based on textbook formulas describing two different regimes: 1) the binary inspiraling analysis treated in Landau and Lifshitz textbook, 2) the plunge of a particle into a black-hole, treated in the Rees-Ruffini-Wheeler textbook as well as 3) the transition between these two regimes following Detweiler's treatment of a particle infalling with non-zero angular momentum onto a black-hole. It is stressed that in order to infer any astrophysical information on the masses of the system both regimes have to be independently and observationally constrained by LIGO, which does not appear to be the case.
\end{abstract}


\maketitle

\section{Introduction}\label{sec:1}

In a recent letter \cite{2016arXiv160504767R}, we have pointed out some issues related to GW 150914, the recently announced gravitational-wave signal by the Advanced Laser Interferometer Gravitational-Wave Observatory (LIGO) \cite{2016PhRvL.116f1102A}. We give in this work additional details.

By using numerical relativity templates of the gravitational-wave emission of black-hole binary mergers, the LIGO-Collaboration concluded that the signal, which lasts 0.2~s with increasing gravitational-wave frequency from 35 to $\sim 150$~Hz, was emitted during the inspiral and merger of a black-hole binary system, followed by the subsequent ringdown of the newly formed black-hole. The binary black-hole parameters obtained from this analysis are: $m_{\rm BH,1} = 36^{+5}_{-4}~M_\odot$, $m_{\rm BH,2} = 29^{+4}_{-4}~M_\odot$, and a luminosity distance to the source $d_L = 410^{+160}_{-180}$~Mpc (cosmological redshift $z=0.09^{+0.03}_{-0.04}$), adopting a flat $\Lambda$-cold-dark-matter cosmology with Hubble parameter $H_0=67.9$~km~s$^{-1}$~Mpc$^{-1}$, and matter and dark-energy density parameter $\Omega_m = 0.306$ and $\Omega_\Lambda = 0.694$, respectively. The mass and spin parameter of the newly formed black-hole are $m_{\rm BH}=62^{+4}_{-4}~M_\odot$ and $\alpha\equiv c J_{\rm BH}/(G m_{\rm BH}^2) = 0.67^{+0.05}_{-0.07}$, respectively, being $J_{\rm BH}$ the black-hole angular momentum.

There are two markedly different regimes in the evolution of coalescing binary black-holes: I) the quasi-circular inspiraling phase and II) the final plunge, merger and subsequent ringdown of the newly formed black-hole. We show in this work that the general features of this system can be inferred from a simple analysis of the gravitational-wave signal on the light of the foundations of the gravitational-wave theory and from three classic results: 1) the analysis of the inspiraling phase of a binary system of point-like particles (regime I, see \cite{1975ctf..book.....L}), 2) the plunge, merger and ringdown analysis based on the approximation of a test-particle falling into a black-hole (regime II, see \cite{1971PhRvL..27.1466D,1972PhRvD...5.2932D} as well as \cite{1974bhgw.book.....R} and appendices therein), and 3) the transition from regime I to regime II based on the approximation of a test-particle falling with non-zero angular momentum into a black-hole \cite{1978ApJ...225..687D,1979ApJ...231..211D}. Our treatment evidences that, for the validity of the binary interpretation, it is essential that both regime I and regime II be independently constrained by observations with comparable accuracy.

\section{Regime I: The inspiral phase}\label{sec:2}

The presence of a binary system is the first fundamental assumption to be confirmed by observational data. A binary system, by emitting gravitational-waves, evolves first through what we call here regime I, namely the inspiral phase in which the binary follows quasi-circular orbits. The gravitational-wave energy spectrum in this phase can be estimated from the traditional formula of the quadrupole emission within the classic point-like approximation \cite{1963PhRv..131..435P,1964PhRv..136.1224P,1974bhgw.book.....R,1975ctf..book.....L}
\begin{equation}\label{eq:dEdfinspiral}
\frac{dE}{df} = \frac{1}{3}(\pi G)^{2/3} \nu M^{5/3} f^{-1/3},
\end{equation}
where $\nu \equiv \mu/M$ is the so-called symmetric mass-ratio parameter, with $\mu=m_1 m_2/M$ the binary reduced mass, $M=m_1+m_2$ the total binary mass, and $f$ is the gravitational-wave frequency. We recall that $f = 2 f_{\rm orb}$, where $f_{\rm orb}$ is the orbital frequency, i.e.~$f_{\rm orb} = \omega_{\rm orb}/(2\pi) = \sqrt{G M/r^3}/(2\pi)$ and $r$ the binary separation distance. We recall that the quantity $M_{\rm chirp}\equiv \nu^{3/5} M$ is referred to in the literature as the binary chirp mass.

\subsection{Analysis of GW 150914}\label{res1}

In order to extract information from the signal related to the regime I it is necessary to make an analysis of the frequency evolution with time since, within this approximation, the binary chirp mass is given by:
\begin{equation}
M_{\rm chirp} = \frac{c^3}{G}\left( \frac{5}{96 \pi^{8/3}} \frac{\dot{f}}{f^{11/3}} \right)^{3/5}.
\end{equation}
We fit the evolution of the frequency with time in GW 150914 which leads to $M^{\rm obs}_{\rm chirp}\approx 30.5~M_\odot$, in agreement with the LIGO analysis, $M^{\rm obs}_{\rm chirp} =30.2^{+2.5}_{-1.9}~M_\odot$, in the detector-frame \cite{2016arXiv160203840T}. We recall that the total mass in the source-frame is $M=M_{\rm obs}/(1+z)$.

From the definition of chirp mass it follows that the total mass of the system is $M = M_{\rm chirp}/\nu^{3/5}$. Thus, since $0<\nu\leq 1/4$, the above value of the chirp mass implies for the total mass a range $70.07\lesssim M_{\rm obs}/M_\odot<\infty$ in the detector-frame.

When the conservative quasi-circular dynamics following the classic point-like approximation breaks down, the regime II, namely the final plunge, merger and ringdown of the newly formed object, sets in (see section~\ref{sec:3}). We denote the gravitational-wave frequency at which the quasi-circular evolution ends as the plunge starting frequency, $f_{\rm plunge}$. We adopt here as an estimate that of the last stable orbit (LSO) of a test-particle around a Schwarzschild black-hole:
\begin{equation}\label{eq:fLSO}
f_{\rm plunge}\approx f_{\rm LSO}=\frac{c^3}{G} \frac{1}{6^{3/2} \pi M} \approx 4.4 \frac{M_\odot}{M}~{\rm kHz}.
\end{equation}
Thus, the total energy radiated during regime I can be estimated from the binding energy of the LSO which, by extrapolation to the case of a binary of comparable masses, reads 
\begin{equation}\label{eq:Einsp}
\Delta E_{\rm inspiral} = (1-\sqrt{8/9}) \mu c^2 = (1-\sqrt{8/9})\nu^{2/5}M_{\rm chirp} c^2,
\end{equation}
which for the above value of the chirp mass implies a range $0<\Delta E_{\rm inspiral}\lesssim M_\odot c^2/(1+z)$.

\section{Regime II: Plunge, merger and ringdown}\label{sec:3}

After the regime I of quasi-circular inspiral evolution, it starts the regime II composed by the plunge, merger and ringdown, as first analyzed in Ref.~\cite{1971PhRvL..27.1466D,1972PhRvD...5.2932D} for a test-particle falling radially into a Schwarzschild black-hole. It was shown in Refs.~\cite{1971PhRvL..27.1466D,1972PhRvD...5.2932D} that the gravitational-wave spectrum in this regime II is dominated by the $l=2$ multipole (quadrupole) emission and that the largest gravitational-wave emission occurs from $r\approx 3 G M/c^2$, at the maximum of the effective potential
%
\begin{align}
&V_l(r) = \left( 1- \frac{2 m_{\rm BH}}{r} \right) \times \\
&\left[\frac{2 \lambda^2 (\lambda+1) r^3 + 6 \lambda^2 m_{\rm BH} r^2 + 18 \lambda m_{\rm BH}^2 r + 18 m_{\rm BH}^3}{r^3 (\lambda r + 3 m_{\rm BH})^2} \right]\nonumber
\end{align}
%
where $\lambda = (l-1)(l+2)/2$, and the black-hole horizon. It was there shown that in the limit of large $l$, the contribution of each multipole to the spectrum peaks at the gravitational-wave frequency
\begin{equation}\label{eq:fpeak}
f^l_{\rm peak}= \frac{c^3}{G}\sqrt{(V_l)_{\rm max}}\approx \frac{c^3}{G}\frac{l}{2 \pi \sqrt{27} M},
\end{equation}
while the total spectrum obtained by summing over all the multipoles peaks at
\begin{equation}\label{eq:fpeak2}
f_{\rm peak} \approx \frac{c^3}{G}\frac{0.05}{M} \approx 10.36 \frac{M_\odot}{M}~{\rm kHz}.
\end{equation}

\begin{figure}
\includegraphics[width=\hsize,clip]{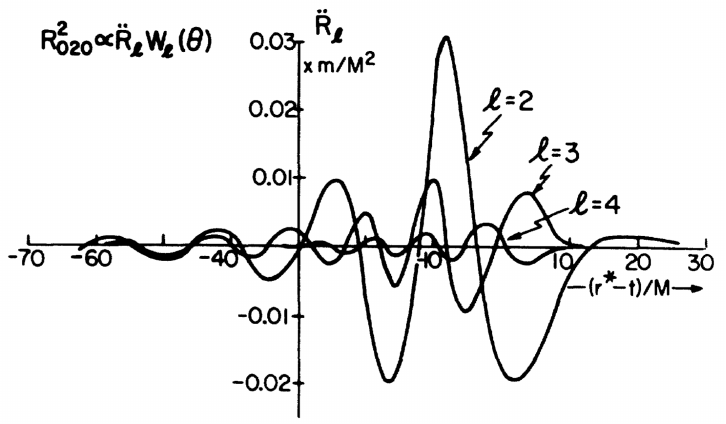}
\caption{$\ddot{R}_l$ factors of the Riemann tensor components as a function of the retarded time for $l = 2$, $3$, $4$. Figure reproduced from Ref.~\cite{1972PhRvD...5.2932D} with kind permission of the authors.}\label{fig:multipoles}
\end{figure}

The signature of the particle infalling is imprinted in the multipolar structure of the signal. The asymptotic expression of the tide-producing components of the Riemann tensor are \cite{1972PhRvD...5.2932D}: $R^2_{020}(r^*,t)=\sum_{l} \ddot{R}_l (r^*,t) W_l(\theta)/(\sqrt[2\pi] 2 r)$, where the angular dependence factor is $W_l(\theta)=(\partial^2/\partial\theta^2-\cot\theta \partial/\partial\theta)Y_{l0}(\theta)$, being $Y_{l 0}$ the spherical harmonics and $r^*=r+2 G M/c^2 \ln[c^2 r/(2 GM)-1]$ (see figure~\ref{fig:multipoles}).

That the multipole spectra obtained in \cite{1971PhRvL..27.1466D} were associated with the $2^l$-pole normal-mode vibrations of the black-hole, excited by the gravitational-wave train produced by the in-falling body, was then shown in \cite{1971ApJ...170L.105P}. Thus, the gravitational-wave spectrum from the peak on is governed by the emission of the vibrational energy of the black-hole driven by gravitational-wave radiation. Such vibrations are known today as black-hole ``ringdown'' or ``ringing tail'' \cite{1971PhRvL..27.1466D}.

\begin{figure}
\includegraphics[width=\hsize,clip]{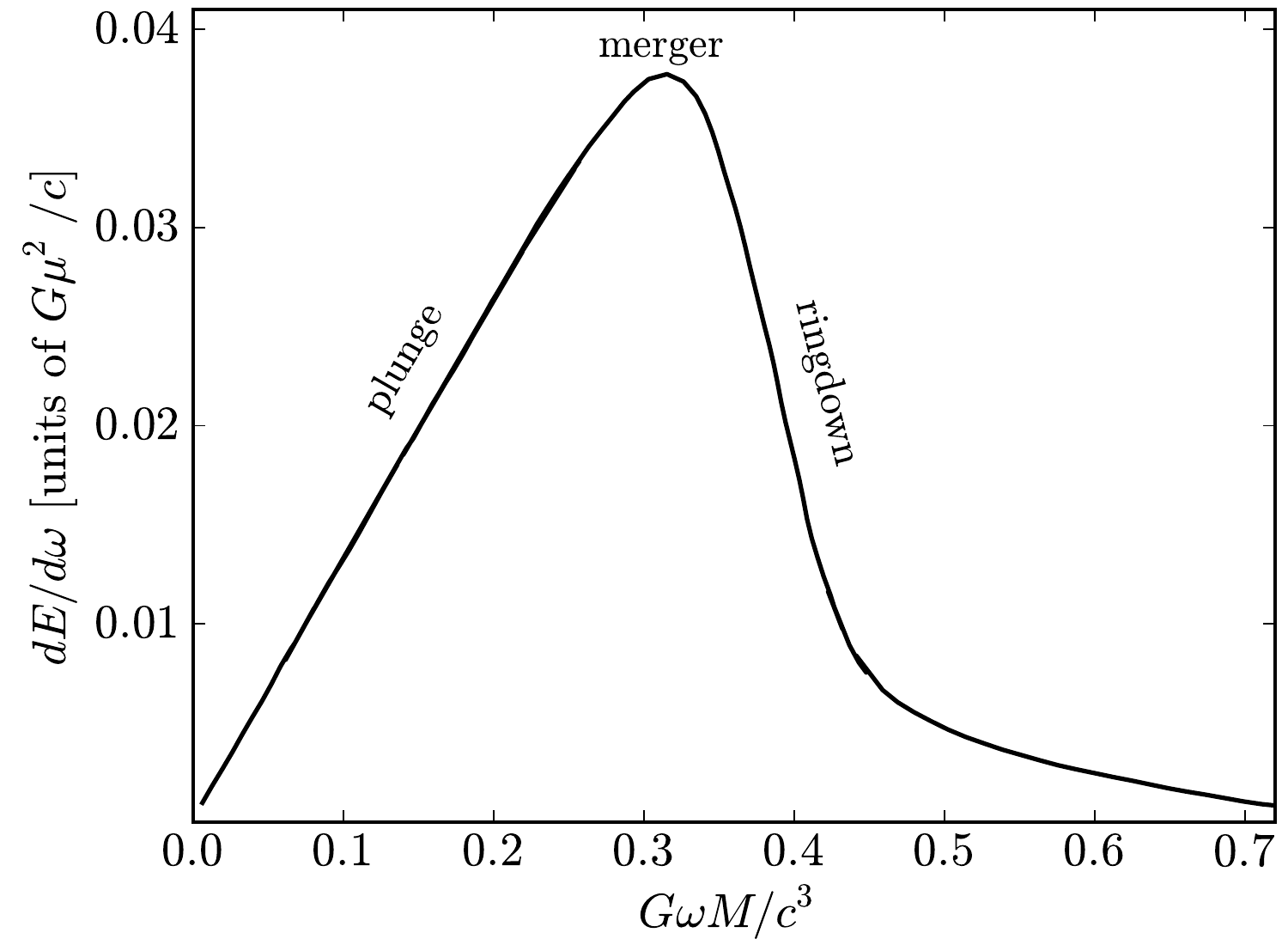}
\caption{Spectrum of the gravitational-wave radiation ($l=2$ multipole) emitted by a test-particle of mass $m$ falling radially into a black-hole of mass $M$ (in geometrical units). This figure has been adapted from the original one in \cite{1971PhRvL..27.1466D}.}\label{fig:DRPP}
\end{figure}

The total gravitational-wave spectrum in the regime II has a peaked form \cite{1971PhRvL..27.1466D,1974bhgw.book.....R} (see figure~\ref{fig:DRPP}): it first has a raising part that follows a power-law behavior, then reaches a maximum to then falling off rapidly. This form of the spectrum was already evident from the first idealized analysis of the radially infalling particle problem in Ref.~\cite{RuffiniWheeler71} that considered the motion of the test-particle in the Schwarzschild metric, but the radiation was there estimated in a flat-space linearized theory of gravity. 

The spectrum raises during the plunge following approximately a power-law
\begin{equation}\label{eq:dEdfmerger2}
\left(\frac{dE}{df}\right)_{\rm plunge} \propto \frac{G \mu^2}{c} \left(\frac{4\pi G M f}{c^3}\right)^{4/3};
\end{equation}
then it reaches a maximum at the peak frequency (\ref{eq:fpeak2}), the approximate point of the merger, and then it falls off at large frequencies approximately as the exponential
\begin{equation}\label{eq:dEdfmerger}
\left(\frac{dE}{df}\right)_{\rm ringdown} \propto \frac{G \mu^2}{c} \exp(-9.9 \times 2\pi G M f/c^3)
\end{equation}
during the ringing. The spectrum of the $l=2$ multipole radiation, obtained numerically in \cite{1971PhRvL..27.1466D}, is shown in figure~\ref{fig:DRPP}. We have indicated the location of the plunge, merger and ringdown phases in the frequency-domain. Clearly, one can obtain an approximate formula of the spectrum from the interpolation function
\begin{equation}
\frac{dE}{df} \approx \left[\frac{1}{(dE/df)_{\rm plunge}} +\frac{1}{(dE/df)_{\rm ringdown}}\right]^{-1}.
\end{equation}

The contribution of each multipole to the total energy radiated during the regime II (plunge+merger+ringdown phase) is $E_{2^l-\rm pole}\approx 0.44 (\mu^2/M) c^2 e^{-2 l}$, so the total energy radiated to infinity in gravitational-waves is \cite{1971PhRvL..27.1466D}:
\begin{equation}\label{eq:Emerger}
\Delta E_{\rm merger} = \sum_{l\geq 2} \int df\left(\frac{dE}{df}\right)_{2^l-{\rm pole}} \approx 0.01 \frac{\mu^2}{M} c^2.
\end{equation}

From all the above we can extract three important theorems:

\begin{enumerate}
\item the final gravitational-wave frequency of the inspiral phase, $f_{\rm LSO}$, is lower than the peak frequency, $f_{\rm peak}$;
\item the energy emitted in gravitational-waves during the total inspiral phase is larger than the energy emitted in the final plunge-merger-ringdown phase;
\item the merger point can be set as the point where the gravitational-wave spectrum, $dE/df$, reaches the maximum value. Then,
\begin{equation}\label{eq:fmerger}
f_{\rm merger} \equiv f_{\rm peak},
\end{equation}
where $f_{\rm peak}$ is given by equation~(\ref{eq:fpeak}).

\end{enumerate}

\subsection{Angular momentum in the merger phase}\label{sec:3c}

It was shown in Ref.~\cite{1978ApJ...225..687D,1979ApJ...231..211D} that the energy emitted during the plunge of a test-particle into a black-hole is affected by the initial angular momentum of the particle. The total energy output in form of gravitational-wave radiation was in Ref.~\cite{1979ApJ...231..211D} computed for selected initial angular momenta of the particle (which correspond to start the plunge of the particle from different orbits). It is worth to notice that in the limit $J=0$, the total energy emitted approaches the numerical value obtained by \cite{1971PhRvL..27.1466D} [see equation~(\ref{eq:Emerger})], namely the one of a radially infalling particle, as expected.

The results of the numerical integration of \cite{1979ApJ...231..211D} are well-fitted (with a maximum error of $\sim 10\%$) by the phenomenological function
\begin{equation}\label{eq:EmergerJ}
\Delta E_{\rm merger} \approx \Delta E_{\rm merger}^{J=0} [1+0.11 \exp(1.53 j)],
\end{equation}
where $j\equiv c J/(G\mu M)$ and $\Delta E_{\rm merger}^{J=0}$ is the energy radiated by a particle falling radially given by equation~(\ref{eq:Emerger}).

Thus, from the knowledge of the angular momentum at the LSO, we can infer the amount of energy emitted during the final merger phase. The energy loss during the regime II is therefore $\Delta E_{\rm merger} \approx 0.24 (\mu^2/M) c^2$, where we have used equations~(\ref{eq:Emerger}), (\ref{eq:EmergerJ}), and the fact that $j_{\rm LSO} = c J_{\rm LSO}/(G \mu M) = 2\sqrt{3}$ is the dimensionless angular momentum of a test-particle in the LSO around a Schwarzschild black-hole. 

From the amount of energy emitted in this final plunge phase, $\Delta E_{\rm merger}$ given by equation~(\ref{eq:EmergerJ}), we can estimate the angular momentum loss by the gravitational-wave emission in the final plunging, $\Delta J_{\rm merger}$, as
\begin{equation}
\Delta J_{\rm merger} \approx \frac{\Delta E_{\rm merger}}{\pi f_{\rm LSO}},
\end{equation}
which leads to $\Delta J_{\rm merger}\approx 3.81 G \mu^2/c$.

\subsection{Analysis of GW 1509014}

\begin{figure}
\includegraphics[width=\hsize,clip]{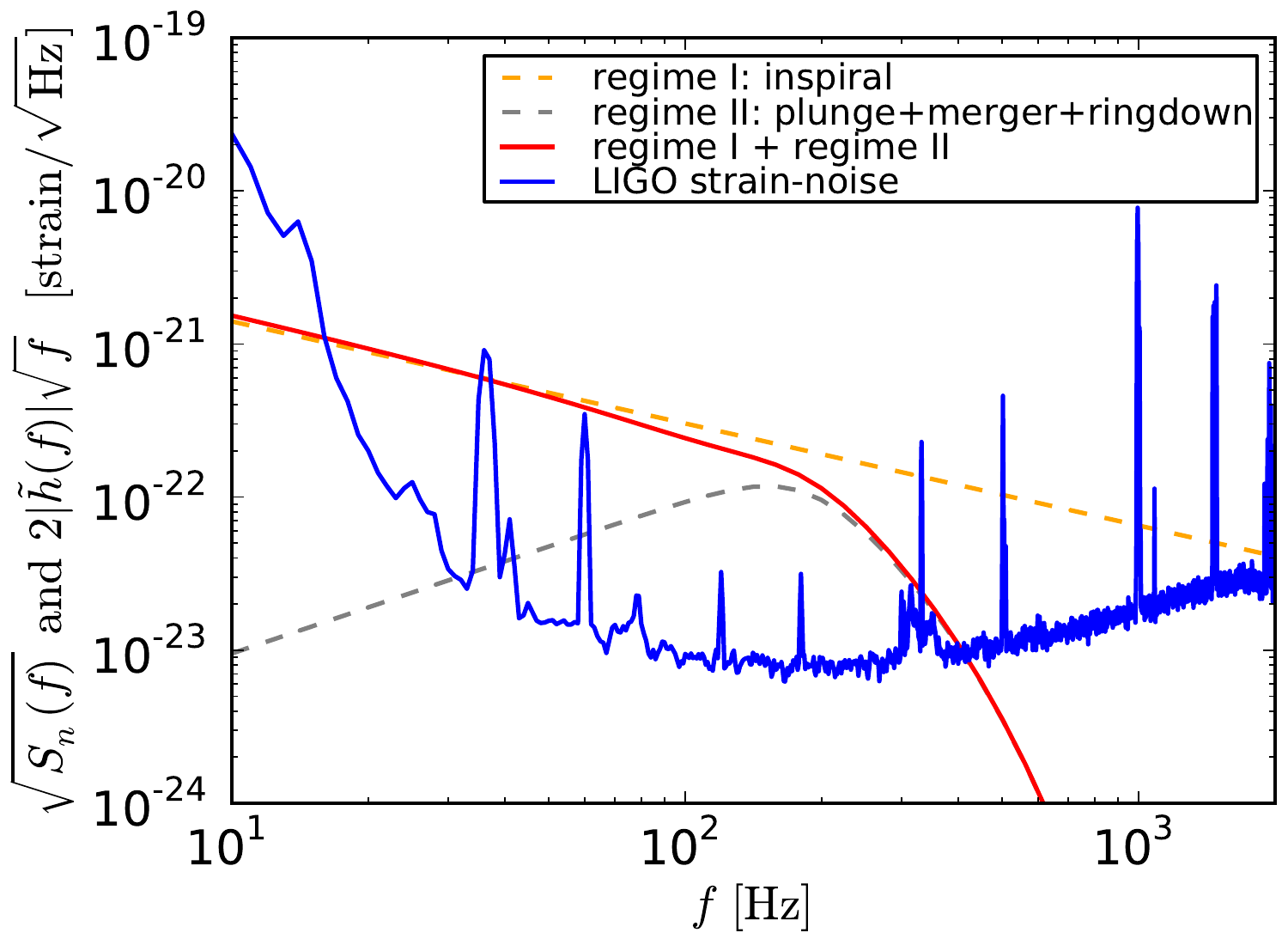}
\caption{Comparison of the amplitude spectral density of the signal in the regimes I (dashed orange curve), II (gray dashed curve) and the total signal (I+II, solid red curve) with the one of the LIGO H1 noise (solid blue curve).}\label{fig:hc}
\end{figure}

It is unfortunate that the signal around 150~Hz occurs just at the limit of the sensitivity of LIGO, not allowing a definite observational characterization of regime II (see figures~\ref{fig:hc}--\ref{fig:rhodif}). Even so, we can proceed with our theoretical analysis by assuming that the evolution of the system follows the above theoretical plunge scenario, originating in a binary system.

We estimate a frequency at maximum strain $f^{\rm obs}_{\rm peak}=144\pm 4$~Hz where the uncertainties are due to the resolution of the discrete Fourier transform used to obtain the spectrogram. By using the aforementioned theorem 3, we can estimate the total mass of the binary equating the observed peak-frequency to the theoretical prediction (\ref{eq:fpeak2}). We obtain the total mass in the detector frame:
\begin{equation}\label{eq:Mth}
M_{\rm obs} = \frac{10.36~{\rm kHz}}{f^{\rm obs}_{\rm peak}} M_\odot \approx 72\pm 2~M_\odot,
\end{equation}
where the quoted errors are associated with the errors in the determination of $f^{\rm obs}_{\rm peak}$. This value is to be compared with the total mass in the detector-frame obtained from the analysis based on numerical relativity templates, $M_{\rm obs}\approx 70.3^{+5.3}_{-4.8}~M_\odot$ \cite{2016arXiv160203840T}, namely, our estimate is off only within a 2.4\% of error with respect to the full numerical relativity analysis. 

From the knowledge of the chirp mass obtained from regime I, and the total binary mass computed here above, we can extract the mass-ratio of the binary. From the definition of symmetric mass-ratio we have
\begin{equation}
\nu = \frac{\mu}{M} = \left(\frac{M_{\rm chirp}}{M}\right)^{5/3}\approx 0.24\pm 0.01,
\end{equation}
which leads to a mass-ratio 
\begin{equation}
q = \frac{m_2}{m_1} = \frac{4\nu}{\bigl(1+\sqrt{1-4\nu}\bigr)^2} \approx 0.67^{+0.33}_{-0.11},
\end{equation}
which is within the numerical relativity analysis value $q=0.79^{+0.18}_{-0.19}$ \cite{2016arXiv160203840T}. Thus, we obtain individual masses
\begin{eqnarray}
m^{\rm obs}_{\rm BH,1}&=&\frac{M_{\rm obs}}{(1+q)}\approx 43.1^{+4.3}_{-7.9}~M_\odot,\\
m^{\rm obs}_{\rm BH,2}&=&\frac{q}{1+q}M_{\rm obs} \approx 28.9^{+6.3}_{-5.9}~M_\odot,
\label{valmass}
\end{eqnarray}
which agree with the numerical relativity values $m_{\rm BH,1} = 39.4^{+5.5}_{-4.9}~M_\odot$, $m_{\rm BH,2} = 30.9^{+4.8}_{-4.4}~M_\odot$ \cite{2016PhRvL.116f1102A,2016arXiv160203840T}.

With the knowledge of the mass of the binary and the one of the binary components we can now estimate the energy emitted in gravitational-waves using equations~(\ref{eq:Einsp}) and (\ref{eq:EmergerJ}) for the regime I and II, respectively. It turns out that for this nearly mass-symmetric binary the two contributions are almost equal with $\Delta E_{\rm inspiral}\approx\Delta E_{\rm merger}\approx 1~M_\odot$, so a total energy emitted $\Delta E_{\rm tot}\approx 2~M_\odot$.

It is worth to stress that the above will remain mere theoretical speculations, unless the hypothesis of a binary nature of the system is confirmed either by independent astrophysical observations (as, e.g., in the case of binary pulsars whose independent observation in the radio frequencies has allowed tests of the general theory of relativity with unprecedented accuracy \cite{1975ApJ...195L..51H,2005ASPC..328...25W}), or by direct observations of the ringdown phase (see section~\ref{sec:5}).

\section{Mass and spin of the formed black-hole}\label{sec:4}

In order to give an estimate of the newly formed black-hole parameters, we can use both energy and angular momentum conservation. Energy conservation implies a mass of the newly formed black-hole
\begin{align}\label{eq:mBH}
    m_{\rm BH} &\approx M - (1-2\sqrt{2}/3)\nu- \Delta E_{\rm merger}/c^2 \\
    &\approx M \beta(\nu),
\end{align}
where $\beta(\nu) \equiv \left[1-\left(1-2\sqrt{2}/3\right)\nu- 0.24 \nu^2\right]$. Angular momentum conservation leads to
\begin{equation}
J_{\rm BH} = J_{\rm LSO} - \Delta J_{\rm merger},
\end{equation}
which implies a dimensionless angular momentum of the newly formed black-hole
\begin{equation}\label{eq:alphafull}
\alpha \equiv \frac{c J_{\rm BH}}{G m_{\rm BH}^2} \approx \frac{2\sqrt{3}\nu - 3.81 \nu^2}{\beta(\nu)^{2}}.
\end{equation}
%

\subsection{Analysis of GW 1509014}\label{sec:4a}

Putting together the equations~(\ref{valmass}), (\ref{eq:mBH}), (\ref{eq:alphafull}), we obtain a straightforward estimate of the parameters of the final black-hole:
\begin{equation}
m_{\rm BH} \approx 70^{+2}_{-2}~M_\odot,\qquad \alpha=\frac{c J_{\rm BH}}{G m^2_{\rm BH}} \approx 0.65^{+0.02}_{-0.02},
\end{equation}
to be compared with the numerical relativity analysis $m_{\rm BH} = 62^{+4}_{-4}~M_\odot$ and $c J_{\rm BH}/(G m^2_{\rm BH}) = 0.67^{+0.05}_{-0.07}$ \cite{2016PhRvL.116f1102A,2016arXiv160203840T}.

\section{Discussion}\label{sec:5}

There are two markedly different regimes in the evolution of coalescing binary black-holes: I) the inspiraling phase; II) the plunge and merger followed by the ringdown phase of the newly formed black-hole. These two regimes have to be constrained by observations with comparable accuracy. In such a case, our analysis shows that it is possible to extract the parameters of the system, indicated in \cite{2016PhRvL.116f1102A}, but from a much simpler analysis of the two regimes in the test-particle approximation without the need of sophisticated numerical simulations. This is quite striking since we would expect that in the real world our test-particle approximation should not be valid in nearly mass-symmetric systems like the one proposed in \cite{2016PhRvL.116f1102A} to explain GW 150914. 

The independent observational confirmation of the two regimes and their subsequent matching is indeed essential in order to evaluate:
\begin{enumerate}
\item[1)] the total mass of the system;
\item[2)] the binary nature of the system and the mass of each binary component;
\item[3)] the formation of the black-hole horizon including the multipolar structure of the ringing;
\item[4)] the energy radiated in gravitational-waves.
\end{enumerate}

\begin{figure}
\includegraphics[width=\hsize,clip]{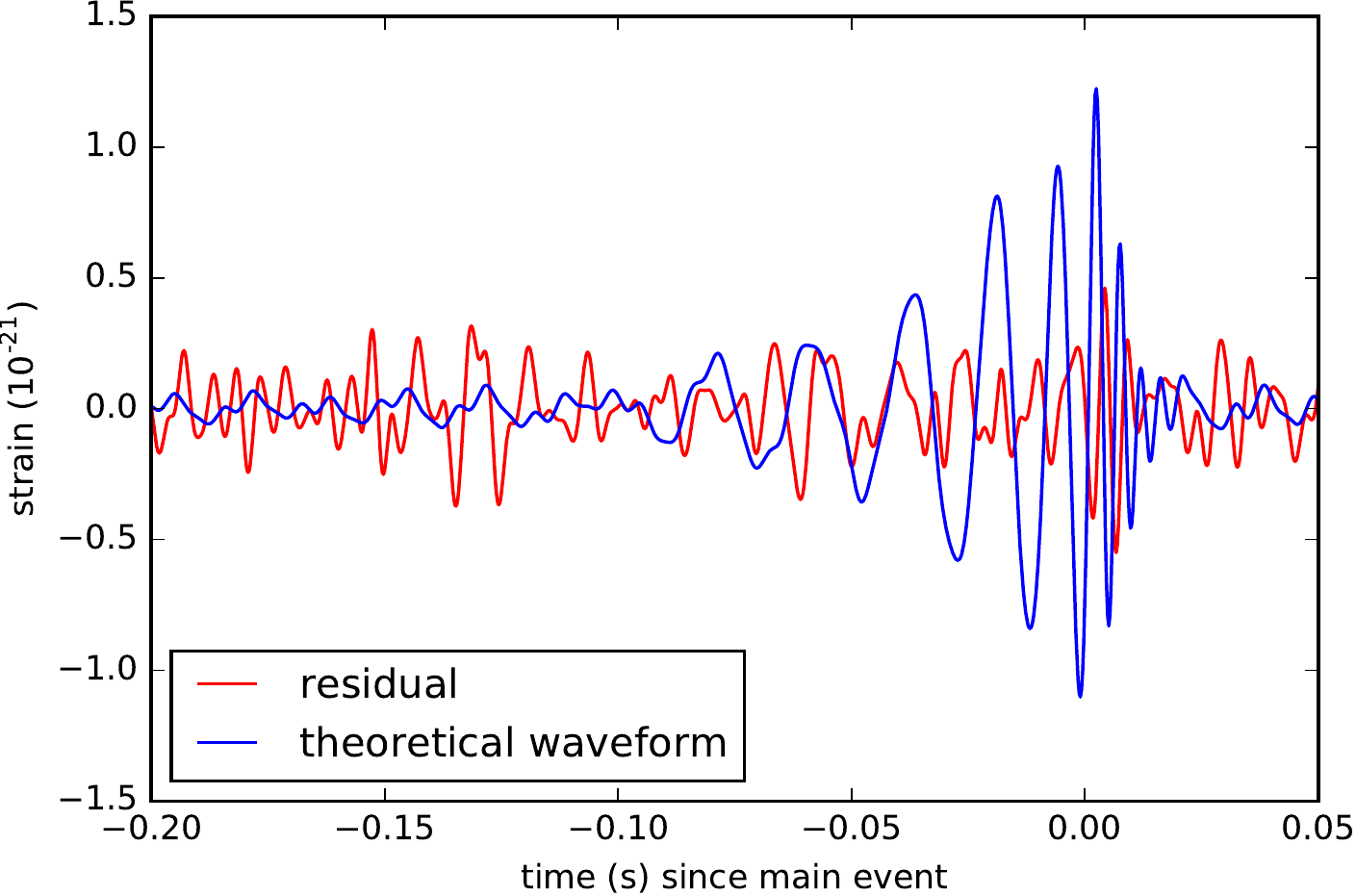}
\caption{(Color online) Blue curve: theoretical waveform of the binary black-hole coalescence used in \cite{2016PhRvL.116f1102A} to explain GW 150914. Red curve: residuals from the subtraction of the theoretical waveform to the filtered H1 detector time series. We can see how the theoretical waveform after the first crest after maximum is of the same order as the residuals, impeding the characterization of the ringdown.}\label{fig:residual}
\end{figure}

\begin{figure}
\includegraphics[width=\hsize,clip]{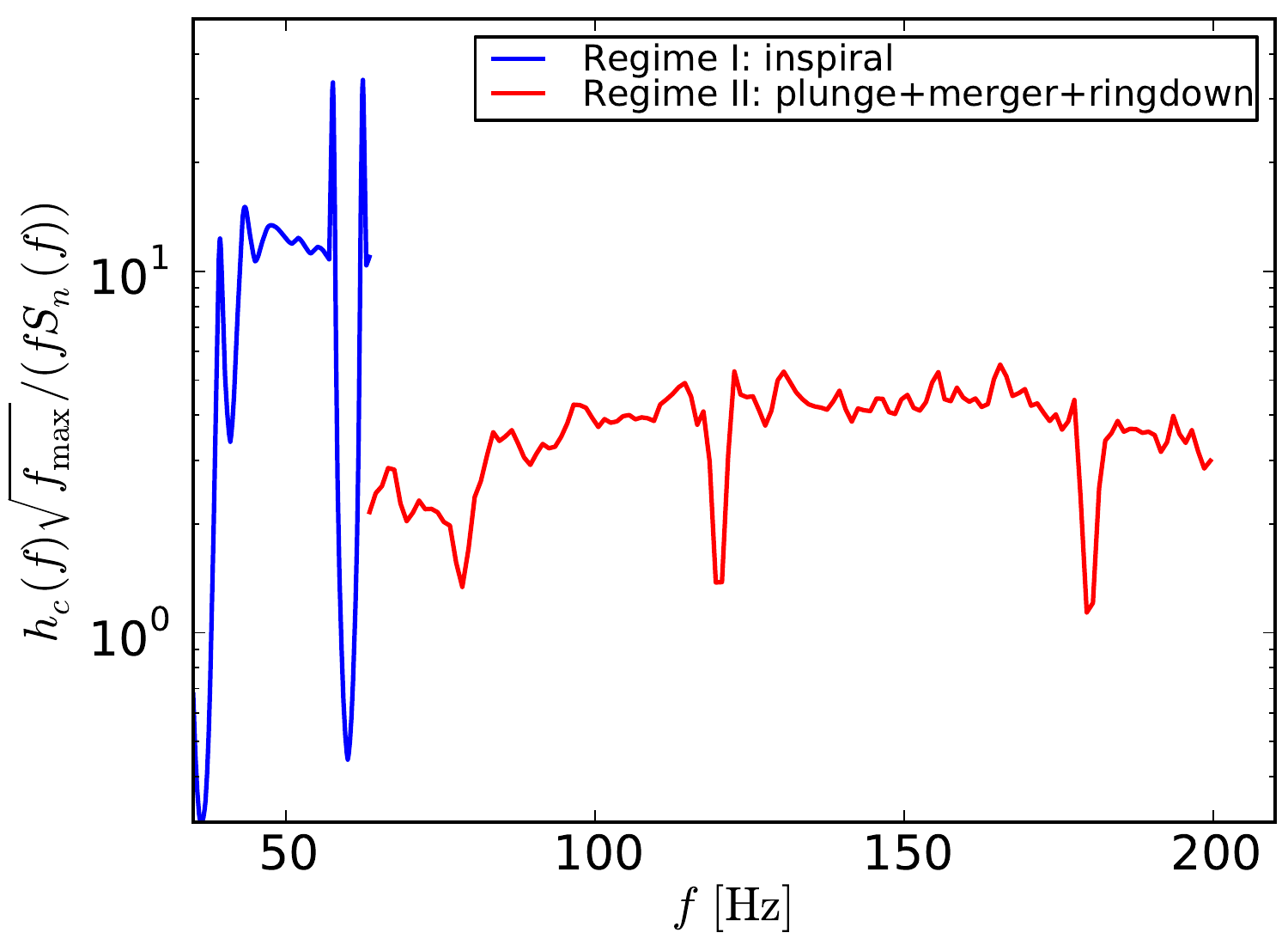}
\caption{(Color online) Estimated signal-to-noise ratio of regimes I (blue curve) and II (red curve) as described in this work. We can see how the signal-to-noise ratio decreases at the end of regime II impeding the characterization of the ringdown. The frequency $f_{\rm max}$ is the maximum observed frequency of the signal in each regime.}\label{fig:rhodif}
\end{figure}

In absence of these verifications no conclusion can be drawn about the nature of the system. It is therefore unfortunate that the signal around 150~Hz occurs just at the limit of the sensitivity of LIGO, not allowing a definite characterization of regime II (see figures~\ref{fig:hc}--\ref{fig:rhodif}). Under these conditions, regime I alone is not sufficient to determine the astrophysical nature of GW 150914, nor to assess that it was produced by a binary black-hole merger leading to a newly formed black-hole.

%

\section{Conclusions}\label{sec:6}

It is by now clear that the LIGO event GW 150914 represent an epochal event just a few weeks after the first run of the advanced LIGO detector and occurring in the most favorable conditions for detectability. The current situation has manifested striking correspondence between our theoretical prediction and the observations. 

The confirmation by observing the late phase of coalescence to a black-hole, the creation of an horizon, the multipolar structure of the ringing (see figures~\ref{fig:hc}--\ref{fig:rhodif}) is now open to further inquire in order to uniquely identify and verify the astrophysical nature of the source. A quest for search of previously discarded signals with lower values of signal-to-noise ratio occurring not in the optimal region of sensitivity nor with the optimal localization and directionality should be addressed. Also important is to focus on the observation of merging processes leading either to a more extended or less extended range of frequencies. 

We can then conclude: 

{\bf A)} if these verifications in past or future observations will materialize, then we are in presence of an outstanding scientific confirmation of the theoretical analysis we here presented on the ground of three fundamental results: 1) the inspiral phase described in \cite{1975ctf..book.....L}, 2) the plunging, merger and ringdown phases described in \cite{1971PhRvL..27.1466D,1972PhRvD...5.2932D} (see also \cite{1974bhgw.book.....R} and appendices therein), and 3) the transition from the quasi-circular inspiral phase to the final merger described in \cite{1979ApJ...231..211D}.

{\bf B)} On the other hand, if these verifications will not succeed, the relevance of the present work will be of service for inquiring on alternative sources not originating from a black-hole created in a black-hole binary merger.

{\bf C)} If this last alternative does not materialize, the relevance of the present work will be more academic and will focus on the perspective that GW 150914 is not of astrophysical nature.

All these three possibilities, at the moment, have to be addressed with equal vigor, each being equally epochal.


\begin{thebibliography}{15}%
\makeatletter
\providecommand \@ifxundefined [1]{%
 \@ifx{#1\undefined}
}%
\providecommand \@ifnum [1]{%
 \ifnum #1\expandafter \@firstoftwo
 \else \expandafter \@secondoftwo
 \fi
}%
\providecommand \@ifx [1]{%
 \ifx #1\expandafter \@firstoftwo
 \else \expandafter \@secondoftwo
 \fi
}%
\providecommand \natexlab [1]{#1}%
\providecommand \enquote  [1]{``#1''}%
\providecommand \bibnamefont  [1]{#1}%
\providecommand \bibfnamefont [1]{#1}%
\providecommand \citenamefont [1]{#1}%
\providecommand \href@noop [0]{\@secondoftwo}%
\providecommand \href [0]{\begingroup \@sanitize@url \@href}%
\providecommand \@href[1]{\@@startlink{#1}\@@href}%
\providecommand \@@href[1]{\endgroup#1\@@endlink}%
\providecommand \@sanitize@url [0]{\catcode `\\12\catcode `\$12\catcode
  `\&12\catcode `\#12\catcode `\^12\catcode `\_12\catcode `\%12\relax}%
\providecommand \@@startlink[1]{}%
\providecommand \@@endlink[0]{}%
\providecommand \url  [0]{\begingroup\@sanitize@url \@url }%
\providecommand \@url [1]{\endgroup\@href {#1}{\urlprefix }}%
\providecommand \urlprefix  [0]{URL }%
\providecommand \Eprint [0]{\href }%
\providecommand \doibase [0]{http://dx.doi.org/}%
\providecommand \selectlanguage [0]{\@gobble}%
\providecommand \bibinfo  [0]{\@secondoftwo}%
\providecommand \bibfield  [0]{\@secondoftwo}%
\providecommand \translation [1]{[#1]}%
\providecommand \BibitemOpen [0]{}%
\providecommand \bibitemStop [0]{}%
\providecommand \bibitemNoStop [0]{.\EOS\space}%
\providecommand \EOS [0]{\spacefactor3000\relax}%
\providecommand \BibitemShut  [1]{\csname bibitem#1\endcsname}%
\let\auto@bib@innerbib\@empty
\bibitem [{\citenamefont {{Rodriguez}}\ \emph {et~al.}(2016)\citenamefont
  {{Rodriguez}}, \citenamefont {{Rueda}},\ and\ \citenamefont
  {{Ruffini}}}]{2016arXiv160504767R}%
  \BibitemOpen
  \bibfield  {author} {\bibinfo {author} {\bibfnamefont {J.~F.}\ \bibnamefont
  {{Rodriguez}}}, \bibinfo {author} {\bibfnamefont {J.~A.}\ \bibnamefont
  {{Rueda}}}, \ and\ \bibinfo {author} {\bibfnamefont {R.}~\bibnamefont
  {{Ruffini}}},\ }\href@noop {} {\bibfield  {journal} {\bibinfo  {journal}
  {submitted to \mnras Lett.; arXiv:1605.04767}\ } (\bibinfo {year} {2016})},\
  \Eprint {http://arxiv.org/abs/1605.04767} {arXiv:1605.04767 [gr-qc]}
  \BibitemShut {NoStop}%
\bibitem [{\citenamefont {{Abbott}}\ \emph {et~al.}(2016)\citenamefont
  {{Abbott}}, \citenamefont {{Abbott}}, \citenamefont {{Abbott}}, \citenamefont
  {{Abernathy}}, \citenamefont {{Acernese}}, \citenamefont {{Ackley}},
  \citenamefont {{Adams}}, \citenamefont {{Adams}}, \citenamefont {{Addesso}},
  \citenamefont {{Adhikari}},\ and\ \citenamefont
  {et~al.}}]{2016PhRvL.116f1102A}%
  \BibitemOpen
  \bibfield  {author} {\bibinfo {author} {\bibfnamefont {B.~P.}\ \bibnamefont
  {{Abbott}}}, \bibinfo {author} {\bibfnamefont {R.}~\bibnamefont {{Abbott}}},
  \bibinfo {author} {\bibfnamefont {T.~D.}\ \bibnamefont {{Abbott}}}, \bibinfo
  {author} {\bibfnamefont {M.~R.}\ \bibnamefont {{Abernathy}}}, \bibinfo
  {author} {\bibfnamefont {F.}~\bibnamefont {{Acernese}}}, \bibinfo {author}
  {\bibfnamefont {K.}~\bibnamefont {{Ackley}}}, \bibinfo {author}
  {\bibfnamefont {C.}~\bibnamefont {{Adams}}}, \bibinfo {author} {\bibfnamefont
  {T.}~\bibnamefont {{Adams}}}, \bibinfo {author} {\bibfnamefont
  {P.}~\bibnamefont {{Addesso}}}, \bibinfo {author} {\bibfnamefont {R.~X.}\
  \bibnamefont {{Adhikari}}}, \ and\ \bibinfo {author} {\bibnamefont
  {et~al.}},\ }\href {\doibase 10.1103/PhysRevLett.116.061102} {\bibfield
  {journal} {\bibinfo  {journal} {\prl}\ }\textbf {\bibinfo {volume} {116}},\
  \bibinfo {eid} {061102} (\bibinfo {year} {2016})},\ \Eprint
  {http://arxiv.org/abs/1602.03837} {arXiv:1602.03837 [gr-qc]} \BibitemShut
  {NoStop}%
\bibitem [{\citenamefont {{Landau}}\ and\ \citenamefont
  {{Lifshitz}}(1975)}]{1975ctf..book.....L}%
  \BibitemOpen
  \bibfield  {author} {\bibinfo {author} {\bibfnamefont {L.~D.}\ \bibnamefont
  {{Landau}}}\ and\ \bibinfo {author} {\bibfnamefont {E.~M.}\ \bibnamefont
  {{Lifshitz}}},\ }\href@noop {} {\emph {\bibinfo {title} {Course of
  theoretical physics - Pergamon International Library of Science, Technology,
  Engineering and Social Studies, Oxford: Pergamon Press, 1975, 4th
  rev.engl.ed.}}}\ (\bibinfo {year} {1975})\BibitemShut {NoStop}%
\bibitem [{\citenamefont {{Davis}}\ \emph {et~al.}(1971)\citenamefont
  {{Davis}}, \citenamefont {{Ruffini}}, \citenamefont {{Press}},\ and\
  \citenamefont {{Price}}}]{1971PhRvL..27.1466D}%
  \BibitemOpen
  \bibfield  {author} {\bibinfo {author} {\bibfnamefont {M.}~\bibnamefont
  {{Davis}}}, \bibinfo {author} {\bibfnamefont {R.}~\bibnamefont {{Ruffini}}},
  \bibinfo {author} {\bibfnamefont {W.~H.}\ \bibnamefont {{Press}}}, \ and\
  \bibinfo {author} {\bibfnamefont {R.~H.}\ \bibnamefont {{Price}}},\ }\href
  {\doibase 10.1103/PhysRevLett.27.1466} {\bibfield  {journal} {\bibinfo
  {journal} {Physical Review Letters}\ }\textbf {\bibinfo {volume} {27}},\
  \bibinfo {pages} {1466} (\bibinfo {year} {1971})}\BibitemShut {NoStop}%
\bibitem [{\citenamefont {{Davis}}\ \emph {et~al.}(1972)\citenamefont
  {{Davis}}, \citenamefont {{Ruffini}},\ and\ \citenamefont
  {{Tiomno}}}]{1972PhRvD...5.2932D}%
  \BibitemOpen
  \bibfield  {author} {\bibinfo {author} {\bibfnamefont {M.}~\bibnamefont
  {{Davis}}}, \bibinfo {author} {\bibfnamefont {R.}~\bibnamefont {{Ruffini}}},
  \ and\ \bibinfo {author} {\bibfnamefont {J.}~\bibnamefont {{Tiomno}}},\
  }\href {\doibase 10.1103/PhysRevD.5.2932} {\bibfield  {journal} {\bibinfo
  {journal} {\prd}\ }\textbf {\bibinfo {volume} {5}},\ \bibinfo {pages} {2932}
  (\bibinfo {year} {1972})}\BibitemShut {NoStop}%
\bibitem [{\citenamefont {{Rees}}\ \emph {et~al.}(1974)\citenamefont {{Rees}},
  \citenamefont {{Ruffini}},\ and\ \citenamefont
  {{Wheeler}}}]{1974bhgw.book.....R}%
  \BibitemOpen
  \bibfield  {author} {\bibinfo {author} {\bibfnamefont {M.}~\bibnamefont
  {{Rees}}}, \bibinfo {author} {\bibfnamefont {R.}~\bibnamefont {{Ruffini}}}, \
  and\ \bibinfo {author} {\bibfnamefont {J.~A.}\ \bibnamefont {{Wheeler}}},\
  }\href@noop {} {\emph {\bibinfo {title} {{Black holes, gravitational waves
  and cosmology}}}}\ (\bibinfo  {publisher} {New York: Gordon and Breach
  Science Publishers Inc.},\ \bibinfo {year} {1974})\BibitemShut {NoStop}%
\bibitem [{\citenamefont {{Detweiler}}(1978)}]{1978ApJ...225..687D}%
  \BibitemOpen
  \bibfield  {author} {\bibinfo {author} {\bibfnamefont {S.~L.}\ \bibnamefont
  {{Detweiler}}},\ }\href {\doibase 10.1086/156529} {\bibfield  {journal}
  {\bibinfo  {journal} {\apj}\ }\textbf {\bibinfo {volume} {225}},\ \bibinfo
  {pages} {687} (\bibinfo {year} {1978})}\BibitemShut {NoStop}%
\bibitem [{\citenamefont {{Detweiler}}\ and\ \citenamefont
  {{Szedenits}}(1979)}]{1979ApJ...231..211D}%
  \BibitemOpen
  \bibfield  {author} {\bibinfo {author} {\bibfnamefont {S.~L.}\ \bibnamefont
  {{Detweiler}}}\ and\ \bibinfo {author} {\bibfnamefont {E.}~\bibnamefont
  {{Szedenits}}, \bibfnamefont {Jr.}},\ }\href {\doibase 10.1086/157182}
  {\bibfield  {journal} {\bibinfo  {journal} {\apj}\ }\textbf {\bibinfo
  {volume} {231}},\ \bibinfo {pages} {211} (\bibinfo {year}
  {1979})}\BibitemShut {NoStop}%
\bibitem [{\citenamefont {{Peters}}\ and\ \citenamefont
  {{Mathews}}(1963)}]{1963PhRv..131..435P}%
  \BibitemOpen
  \bibfield  {author} {\bibinfo {author} {\bibfnamefont {P.~C.}\ \bibnamefont
  {{Peters}}}\ and\ \bibinfo {author} {\bibfnamefont {J.}~\bibnamefont
  {{Mathews}}},\ }\href {\doibase 10.1103/PhysRev.131.435} {\bibfield
  {journal} {\bibinfo  {journal} {Physical Review}\ }\textbf {\bibinfo {volume}
  {131}},\ \bibinfo {pages} {435} (\bibinfo {year} {1963})}\BibitemShut
  {NoStop}%
\bibitem [{\citenamefont {{Peters}}(1964)}]{1964PhRv..136.1224P}%
  \BibitemOpen
  \bibfield  {author} {\bibinfo {author} {\bibfnamefont {P.~C.}\ \bibnamefont
  {{Peters}}},\ }\href {\doibase 10.1103/PhysRev.136.B1224} {\bibfield
  {journal} {\bibinfo  {journal} {Physical Review}\ }\textbf {\bibinfo {volume}
  {136}},\ \bibinfo {pages} {1224} (\bibinfo {year} {1964})}\BibitemShut
  {NoStop}%
\bibitem [{\citenamefont {{The LIGO Scientific Collaboration}}\ and\
  \citenamefont {{the Virgo Collaboration}}(2016)}]{2016arXiv160203840T}%
  \BibitemOpen
  \bibfield  {author} {\bibinfo {author} {\bibnamefont {{The LIGO Scientific
  Collaboration}}}\ and\ \bibinfo {author} {\bibnamefont {{the Virgo
  Collaboration}}},\ }\href@noop {} {\bibfield  {journal} {\bibinfo  {journal}
  {ArXiv e-prints}\ } (\bibinfo {year} {2016})},\ \Eprint
  {http://arxiv.org/abs/1602.03840} {arXiv:1602.03840 [gr-qc]} \BibitemShut
  {NoStop}%
\bibitem [{\citenamefont {{Press}}(1971)}]{1971ApJ...170L.105P}%
  \BibitemOpen
  \bibfield  {author} {\bibinfo {author} {\bibfnamefont {W.~H.}\ \bibnamefont
  {{Press}}},\ }\href {\doibase 10.1086/180849} {\bibfield  {journal} {\bibinfo
   {journal} {\apjl}\ }\textbf {\bibinfo {volume} {170}},\ \bibinfo {pages}
  {L105} (\bibinfo {year} {1971})}\BibitemShut {NoStop}%
\bibitem [{\citenamefont {{Ruffini}}\ and\ \citenamefont
  {{Wheeler}}(1971)}]{RuffiniWheeler71}%
  \BibitemOpen
  \bibfield  {author} {\bibinfo {author} {\bibfnamefont {R.}~\bibnamefont
  {{Ruffini}}}\ and\ \bibinfo {author} {\bibfnamefont {J.~A.}\ \bibnamefont
  {{Wheeler}}},\ }in\ \href@noop {} {\emph {\bibinfo {booktitle} {Proceedings
  of the Cortona Symposium on Weak Interactions}}},\ \bibinfo {editor} {edited
  by\ \bibinfo {editor} {\bibfnamefont {L.}~\bibnamefont {{Radicati}}}}\
  (\bibinfo {year} {1971})\BibitemShut {NoStop}%
\bibitem [{\citenamefont {{Hulse}}\ and\ \citenamefont
  {{Taylor}}(1975)}]{1975ApJ...195L..51H}%
  \BibitemOpen
  \bibfield  {author} {\bibinfo {author} {\bibfnamefont {R.~A.}\ \bibnamefont
  {{Hulse}}}\ and\ \bibinfo {author} {\bibfnamefont {J.~H.}\ \bibnamefont
  {{Taylor}}},\ }\href {\doibase 10.1086/181708} {\bibfield  {journal}
  {\bibinfo  {journal} {\apjl}\ }\textbf {\bibinfo {volume} {195}},\ \bibinfo
  {pages} {L51} (\bibinfo {year} {1975})}\BibitemShut {NoStop}%
\bibitem [{\citenamefont {{Weisberg}}\ and\ \citenamefont
  {{Taylor}}(2005)}]{2005ASPC..328...25W}%
  \BibitemOpen
  \bibfield  {author} {\bibinfo {author} {\bibfnamefont {J.~M.}\ \bibnamefont
  {{Weisberg}}}\ and\ \bibinfo {author} {\bibfnamefont {J.~H.}\ \bibnamefont
  {{Taylor}}},\ }in\ \href@noop {} {\emph {\bibinfo {booktitle} {Binary Radio
  Pulsars}}},\ \bibinfo {series} {Astronomical Society of the Pacific
  Conference Series}, Vol.\ \bibinfo {volume} {328},\ \bibinfo {editor} {edited
  by\ \bibinfo {editor} {\bibfnamefont {F.~A.}\ \bibnamefont {{Rasio}}}\ and\
  \bibinfo {editor} {\bibfnamefont {I.~H.}\ \bibnamefont {{Stairs}}}}\
  (\bibinfo {year} {2005})\ p.~\bibinfo {pages} {25},\ \Eprint
  {http://arxiv.org/abs/astro-ph/0407149} {astro-ph/0407149} \BibitemShut
  {NoStop}%
\end{thebibliography}

%

\end{document}